# A Model of Neutrinos

**Alan Chodos**[*]
*Haseltine Systems, 2181 Jamieson Ave., Suite 1606*
*Alexandria, VA 22314*

**Abstract**: We propose a model for neutrinos based on symmetry under light-cone reflection (LCR), which was introduced in a previous paper[10]. LCR is realized using a minimal substitution, which allows the equations of motion to be solved after a suitable coordinate transformation. Some consequences of this proposal are discussed.

**Introduction**

Representations of the Poincaré group admit 3 classes of momenta: timelike, lightlike and spacelike[1]. So far, however, particles corresponding only to the first two have been found in nature. Some would argue [2] that tachyons (as particles in the third class have come to be called) inescapably lead to violations of causality, and hence can be ruled out by pure thought, irrespective of what experiment has to say on the matter.

If, however, we reject that view as unscientific, we are left, on the one hand, with the question of experimental evidence for or against the existence of tachyons, and, on the other, with theoretical issues such as what a sensible theory of tachyons should look like, and how to construct it.

In this paper, we shall examine, in particular, the possibility that one or more neutrino species might be tachyonic. The reason for this choice is the same that led the author and collaborators to consider this idea some 30 years ago [3]: of all the known elementary particles, the only one that could possibly be a tachyon is the neutrino. If tachyons other than neutrinos exist, they are so exotic, and interact so weakly with ordinary matter that they have thus far eluded detection. It is better to begin under the one lamppost that we know, before undertaking a search in the dark.

On the experimental side, one is living in the long shadow of the claim that was made [4], and later withdrawn [5], by the OPERA collaboration a few years ago. To many it was a great relief that, when the dust settled, the velocity of neutrinos appeared to be consistent with that of light. But if one assumes the dispersion formula for spacelike momenta given by relativity,

$$E^2 = \vec{p}^2 - m^2 \qquad (1)$$

then, using the energy of the neutrinos and their velocity as measured by OPERA, the mass parameter comes out to be a couple of hundred MeV, nowhere near the meV range that one expects for neutrinos. So the withdrawal of the claim by OPERA was a relief not only to skeptics, but also to proponents of a tachyonic neutrino. A time-of-flight measurement would have to be many orders of magnitude more sensitive than that of OPERA to determine the neutrino mass, be its velocity faster or slower than that of light.

[*] chodos@aps.org

Other evidence for tachyonic neutrinos has been put forward by Ehrlich[6], who performs an innovative analysis of astrophysical data.

On the theoretical side, the original work[3] used the simple Lagrangian

$$L = \overline{\psi}( i\gamma_5\gamma^\mu \nabla_\mu - m)\psi \qquad (2)$$

which is formally Hermitian, and which straightforwardly imposes the dispersion formula (1). In addition, the appearance of $\gamma_5$ in the kinetic term hints at a deeper understanding of the chiral nature of the weak interactions. Jentschura and Wundt have thoroughly investigated the solutions to equation (2). [7]

However, despite much effort, a sensible quantum theory based on this Lagrangian has never been constructed[8]. [One has to be careful, though, in using "sensible" to describe a theory of tachyons. There is a list of fundamental properties, including, among others, a Hermitian Lagrangian, a Hamiltonian that is bounded below, and locality, which long experience has classified as "sensible", but which may not be possible to implement *in toto* in a theory of tachyons. One might suspect, for example, that such a theory should be inherently non-local, because tachyons propagate outside the light cone. Another example of tachyons violating conventional expectations was given by the early work Feinberg[9] (who coined the word "tachyon"). He constructed a Lorentz-invariant theory of scalar tachyons, but at the cost of quantizing them with anti-commutators.]. Nor was there a more compelling reason than simplicity for the choice of the Lagrangian (2).

We shall put aside this Lagrangian and instead elaborate an idea proposed in a previous paper[10]. We surrender full Lorentz invariance, and work within the framework of Very Special Relativity (VSR), as proposed[11] by Cohen and Glashow some years ago. Having lost part of one symmetry, we shall impose another, discrete symmetry that we call Light Cone Reflection (LCR). As its name implies, LCR takes timelike four-vectors to spacelike ones, and vice-versa. The hoped-for consequence of this symmetry will be the appearance of tachyons as partners of ordinary particles. We shall see that this is in fact realized, although because we have given up full Lorentz invariance, the dispersion formula is not exactly the usual one.

Because LCR is a spacetime-dependent transformation, to achieve invariance we need to turn the derivatives in the Lagrangian into covariant ones. In reference [10] we introduced a vector field for that purpose. Here we adopt a more economical approach, making use of only an additional scalar field, which furthermore is subject to a constraint. We then reduce the equations of motion to a linear system of differential equations with constant coefficients, which can be solved by Fourier transformation.

**Very Special Relativity**

We employ the Sim(2) version of VSR, which breaks Lorentz invariance through the appearance in the Lagrangian of a null vector $n^\mu$. L must be invariant under those Lorentz transformations that leave $n^\mu$ invariant, up to a possible overall scaling of $n^\mu$ (the scaling corresponds to a boost in the $\vec{n}$ direction). We choose our coordinates such that $n^\mu$ takes the form $n^\mu = (1,0,0,1)$.

With this choice of $n^\mu$, the 4 generators of Sim(2) are

$G_1 = M^{12}$; $G_2 = M^{13} + M^{01}$; $G_3 = M^{23} + M^{02}$; and $G_4 = M^{03}$ where the M's are the familiar generators of the Lorentz group, such that $M^{ij}$ generate rotations and $M^{0i}$ generate boosts.

The wave equation for neutrinos studied by Cohen and Glashow is

$$\left(p_\mu - \frac{m^2}{2n\cdot p}n_\mu\right)\gamma^\mu \psi = 0, \qquad (3)$$

where $\psi$ is the neutrino field. It is easy to see that the spectrum of the solutions is just $p^2 = m^2$. Furthermore, because of the factor $n_\mu \gamma^\mu$, equation (3) is chirally invariant. It can be derived from a Lagrangian that was introduced by Alvarez and Vidal[12]:

$$L = i\overline{\psi}\gamma^\mu \partial_\mu \psi + i\overline{\chi}(n\cdot\partial)\rho + i\overline{\rho}(n\cdot\partial)\chi + \frac{i}{2}m\{\overline{\chi}(n\cdot\gamma)\psi + \overline{\rho}\psi - h.c.\} \qquad (4)$$

The auxiliary fields $\chi$ and $\rho$ avoid the introduction of explicitly non-local terms. The field $\chi$ absorbs the factor under which $n^\mu$ scales, and $\rho$ transforms oppositely to $\psi$ and $\chi$ under chiral transformations.

We shall not discuss the question of the extent to which VSR is consistent with observation. This will depend, of course, on which interactions break Lorentz symmetry down to VSR. If VSR appears only in the neutrino sector, it is much less likely to conflict with experiment than if VSR is used to describe a broader swath of the Standard Model.

**Light Cone Reflection**

Following reference [10], we can make use of the null vector $n^\mu$ to define a transformation

$$x^\mu \rightarrow x'^\mu = x^\mu - \frac{x^2}{n\cdot x}n^\mu. \qquad (5)$$

We see immediately that $x'^2 = -x^2$. If we iterate the transformation, we get back where we started. Essentially the transformation is a mapping between the interior of the light cone and that part of the exterior of the light cone satisfying $n \cdot x \neq 0$. Roughly speaking, it amounts to unzipping the light cone along the n direction, reflecting in the resulting plane, and then zipping it back up again. We call it Light Cone Reflection (LCR). Additional elementary properties were discussed in reference [10].

Our task is to modify the Lagrangian (4) to make it LCR invariant. In addition to VSR and LCR, we shall also insist on preserving chiral symmetry. With one exception, we shall include all possible terms that are simultaneously VSR, LCR and chirally invariant.

Let us write the Lagrangian L as

$$L = L_\psi + L_{\chi\rho} + L_M \qquad (6)$$

in an obvious notation. To make it LCR invariant, we proceed in two steps. First, we note that the derivative $n \cdot \partial$ is odd under LCR. To prevent $L_{\chi\rho}$ from changing sign, we must choose one of ρ or χ to be odd under LCR. We shall choose ρ. Then the ρ-dependent terms in $L_M$ will change sign under LCR, unless we find a way to compensate. We do this by doubling number of ψ fields to a pair that we label $\psi_\pm$, which are even and odd under LCR respectively. We can then couple $\psi_+$ to χ and $\psi_-$ to ρ.

Second, we must replace the ordinary derivatives occurring in $L_\psi$ with covariant ones $D_\mu$ which transform as $D_\mu \to D'_\mu$ under the LCR transformation $x^\mu \to x'^\mu$. In reference [10] we did this with the collusion of a vector field, similar to what one does to implement gauge invariance. This seems excessive, however, since a gauge transformation is continuous whereas LCR is discrete. To find the "minimal substitution", we introduce a scalar field ϕ, and furthermore impose the constraint

$$n \cdot \partial \phi = 1 . \qquad (7)$$

We then define the covariant derivative as

$$D_\mu = \partial_\mu - (\partial_\mu \phi) \, n \cdot \partial \quad . \qquad (8)$$

Under a VSR transformation that takes $n^\mu$ to $\alpha n^\mu$, ϕ transforms as

$$\phi(x) \to (1/\alpha) \phi(x')$$

and under LCR it transforms as

$$\phi(x) \to -\phi(x') \; .$$

These transformations leave the constraint invariant, and also insure that $D_\mu$ indeed transforms covariantly under LCR. In an alternative formulation, we could dispense with the constraint and instead define the covariant derivative as

$$D_\mu = \partial_\mu - (\partial_\mu \phi / n \cdot \partial \phi) \, n \cdot \partial \ .$$

We can then impose the simpler transformation ϕ(x) → ϕ(x') under both VSR and LCR, but at the cost of having an inverse power of n·∂ϕ appear in the Lagrangian. For the rest of this paper we shall stick to the formulation with the constraint. Note that, in either case, we have $n^\mu D_\mu = 0$, and also $D_\mu \phi = 0$.

**Generalized Lagrangian**

In reference [12], a single mass parameter was introduced, because the aim was simply to reproduce equation (3). We generalize this in several ways: first, we allow separate mass parameters for ψ₊ and ψ₋ ; second, we allow these parameters to be complex (making sure the Lagrangian remains Hermitian, of course); and third, we include Majorana mass terms as well as Dirac ones. Bearing in mind the need to maintain VSR, LCR and chiral symmetry, we find the following terms in the Lagrangian:

$$L_\psi = i\overline{\psi}_+ \gamma^\mu D_\mu \psi_+ + i\overline{\psi}_- \gamma^\mu D_\mu \psi_- \ , \qquad (9)$$

$$L_{\chi\rho} = i[\overline{\chi}(n \cdot \partial)\rho + \overline{\rho}(n \cdot \partial)\chi] \ , \qquad (10)$$

$$L_M = i\{M_1 \overline{\chi} \gamma^\mu n_\mu \psi_+ + M_2 \overline{\rho}\psi_- + M_3^* \overline{\psi}_+ \gamma^\mu n_\mu \chi^c + M_4^* \overline{\psi}_- \rho^c - h.c.\} \ . \qquad (11)$$

Here ψᶜ is, as usual, the charge conjugate of ψ, defined by

$$\psi^c = C\overline{\psi}^T \qquad (12)$$

C is the charge conjugation matrix, satisfying $C^{-1}\gamma^\mu C = -(\gamma^\mu)^T$. Under the chiral symmetry, ψ± and χ are multiplied by exp{iαγ₅} while ρ picks up a factor exp{−iαγ₅}, for arbitrary real α.

Given the degrees of freedom above, we have included all possible terms consistent with VSR, LCR, and chiral symmetry, except for a term proportional to $i\overline{\rho}\gamma^\mu D_\mu \rho$. Such a term seems to play no role in solutions to the equations of motion, so we omit it for simplicity in this paper, but it should be borne in mind that it might be needed in future developments.

**Equations of Motion**

The degrees of freedom that we vary to get the equations of motion are ψ±, χ and ρ and their conjugates, as well as the scalar ϕ, subject to the constraint (7). Because the action is Hermitian, the equations obtained by varying the Fermi fields and their conjugates will simply be Hermitian conjugates of each other, so we shall display only those that follow from varying the conjugates. We shall leave discussion of the φ equation to the next section.

The Fermion equations of motion are:

$$\gamma^\mu D_\mu \psi_+ - M_1^* \gamma^\mu n_\mu \chi + M_3^* \gamma^\mu n_\mu \chi^c = 0 ; \tag{13}$$

$$\gamma^\mu D_\mu \psi_- - M_2^* \rho + M_4^* \rho^c = 0 ; \tag{14}$$

$$n^\mu \partial_\mu \rho + M_1 n^\mu \gamma_\mu \psi_+ - M_3^* n^\mu \gamma_\mu \psi_+^c = 0 ; \tag{15}$$

$$n^\mu \partial_\mu \chi + M_2 \psi_- + M_4^* \psi_-^c = 0 . \tag{16}$$

Were it not for the unknown function φ that appears in $D_\mu$, these equations could be solved by Fourier transformation. We can obtain a set of equations independent of φ by means of a judicious choice of coordinates.

Let us first define light-cone coordinates

u = $n_\mu x^\mu$ = t − z   and   v = t + z .

We shall take u to be our evolution parameter. Note that $n^\mu \partial_\mu = 2 \frac{\partial}{\partial v}$.

Because of the constraint (7), φ can be written

$$\phi = \frac{1}{2} v + h(u, \vec{x}_\perp) , \tag{17}$$

where $\vec{x}_\perp = (x, y)$ and h is an arbitrary function of the indicated variables. Define a four-vector $\nabla_{\perp \mu}= (0, \partial_x, \partial_y, 0)$

Then

$$D_\mu = n_\mu (\partial_u - 2\dot{h}\partial_v) + \nabla_{\perp \mu} - (2\nabla_{\perp \mu} h)\partial_v , \tag{18}$$

where the over-dot indicates differentiation with respect to u.

We transform variables as follows:

$$u' = u ; \quad \vec{x}'_\perp = \vec{x}_\perp ; \quad v' = v + 2h . \tag{19}$$

Note, from (17), that we are just replacing v with ϕ. This might strike some as questionable, because we are demoting ϕ from a dynamical degree of freedom to a coordinate. We proceed nonetheless.

The derivatives become:

$$\partial_u = \partial_{u'} + 2\dot{h}\partial_{v'} \;;\; \vec{\nabla} = \vec{\nabla}' + 2(\vec{\nabla}h)\partial_{v'} \;;\; \partial_v = \partial_{v'} \;.$$

Hence the covariant derivative is

$$D_\mu = n_\mu \partial_{u'} + \nabla'_{\perp\mu} \;.$$

We refer to the primed variables as "special coordinates". Using the standard γ-matrix representation

$$\gamma^0 = \begin{pmatrix} I & 0 \\ 0 & -I \end{pmatrix} \;;\; \vec{\gamma} = \begin{pmatrix} 0 & \vec{\sigma} \\ -\vec{\sigma} & 0 \end{pmatrix}$$

we expand

$$\psi_\pm = \sum_{i=1}^{4} f_i^{(\pm)} |b_i>$$

in a spinor basis given by

$$|b_1> = \begin{pmatrix} \uparrow \\ \uparrow \end{pmatrix} \;;\; |b_2> = \begin{pmatrix} \downarrow \\ \downarrow \end{pmatrix} \;;\; |b_3> = \begin{pmatrix} \uparrow \\ -\uparrow \end{pmatrix} \;;\; |b_4> = \begin{pmatrix} \downarrow \\ -\downarrow \end{pmatrix},$$

where ↑ and ↓ denote eigenstates of $\sigma_z$ with eigenvalues +1 and -1 respectively. $|b_1>$ and $|b_4>$ are annihilated by $n \cdot \gamma$. $|b_1>$ and $|b_2>$ are eigenstates of $\gamma_5$ with eigenvalue +1, whereas $|b_3>$ and $|b_4>$ are eigenstates of $\gamma_5$ with eigenvalue −1.

Analyzing the equations of motion (13) – (16) in this basis, we find that $\psi_\pm$ must be independent of the transverse variables $\vec{x}$, depending only on u' and v'. In the original coordinates, $\psi_\pm$ will therefore depend on $\vec{x}$ only through the variable h.

We differentiate (13) and (14) with respect to v', and then use (15) and (16) to obtain

$$2n \cdot \gamma \partial_{u'} \partial_{v'} \psi_+ = n \cdot \gamma [B\psi_- + A^* \psi_-^c] \tag{20}$$

and

$$2n \cdot \gamma \partial_{u'} \partial_{v'} \psi_- = n \cdot \gamma [B^* \psi_+ + A^* \psi_+^c] \;, \tag{21}$$

which leads to the following equations for the f's:

$$\lambda f_2^{(+)} - B f_2^{(-)} - A^* f_3^{(-)*} = 0 \tag{22}$$

$$\lambda f_3^{(+)} - B f_3^{(-)} - A^* f_2^{(+)*} = 0 \tag{23}$$

$$\lambda f_2^{(-)} - B^* f_2^{(+)} - A^* f_3^{(+)*} = 0 \tag{24}$$

$$\lambda f_3^{(-)} - B^* f_3^{(+)} - A^* f_2^{(+)*} = 0 \ . \tag{25}$$

Here $\lambda = 2\partial_{u'}\partial_{v'}$ , and A and B are the combinations

$$A = M_2 M_3 - M_1 M_4 \ \text{and} \ B = M_3^* M_4 - M_1^* M_2 \ .$$

These equations determine four possible eigenvalues for $\lambda$:

$$\lambda = \pm(|A| + |B|) \ \text{and}$$

$$\lambda = \pm(|A| - |B|) \ .$$

Within the confines of the special coordinate system, where there is no dependence on the transverse coordinates, we can think of $\lambda$ as $-\frac{1}{2}\mathcal{M}^2$ , where $\mathcal{M}^2$ represents the mass-squared operator. In this system, the spectrum exhibits the symmetry between positive and negative mass-squared that we anticipated when we introduced LCR.

The equations of motion have reduced to equations for $f_2^{(\pm)}$ and $f_3^{(\pm)}$ . The coefficients $f_1^{(\pm)}$ and $f_4^{(\pm)}$ are undetermined. That is, the part of $\psi_\pm$ annihilated by $n \cdot \gamma$ must be regarded as unphysical. Coupling $\psi_\pm$ to other fields will have to be arranged in such a way that the unphysical part does not participate in the interaction.

Another feature of equations (22) – (25) is that $f_2^{(\pm)}$ couple to the complex conjugates of $f_3^{(\pm)}$ (and of course vice-versa). For a given $\lambda$, the Fourier decomposition of the coefficients can therefore be divided into two cases:

Case I : $f_2^{(\pm)} \sim exp[-i(Eu' - pv')]$ and $f_3^{(\pm)} \sim exp[i(Eu' - pv')]$   or

Case II : $f_2^{(\pm)} \sim exp[i(Eu' - pv')]$ and $f_3^{(\pm)} \sim exp[-i(Eu' - pv')]$ ,

where $E > 0$  and $\lambda = 2Ep$ . These two cases correspond to the existence of particles and anti-particles.

**Variation of ϕ**

In addition to varying the Fermi fields, we must also take into account the equation that follows from the variation of the scalar field ϕ that appears in the covariant derivative (8). Because of the constraint (7), this amounts to varying the field h that is defined in (17). This field, and its variation, are independent of the coordinate v; hence the condition we obtain upon variation of the action will be integrated over v.

The field h is contained in the covariant derivative $D_\mu$, which appears only in $L_\psi$. Invariance of the action under variation of h yields the condition

$$\frac{\partial}{\partial u}\int dv\, [\overline{\psi}_+ n_\mu \gamma^\mu \partial_v \psi_+ + \overline{\psi}_- n_\mu \gamma^\mu \partial_v \psi_-] + \vec{\nabla}_\perp \cdot \int dv\, [\overline{\psi}_+ \vec{\gamma} \partial_v \psi_+ + \overline{\psi}_- \vec{\gamma} \partial_v \psi_-] = 0.$$

(26)

We argue that, for fields obeying the equations of motion, this condition is automatically satisfied, and hence yields no new information.

We have found that when $\psi_\pm$ obey the equations of motion, they are functions of only the two variables u' = u and v' = v + 2h, where h is independent of v but otherwise arbitrary. In the second integral of (24), we can perform the change of variable v → v', thereby removing all dependence on $\vec{x}_\perp$. The second term then vanishes because of the gradient operator.

To see that the first term also vanishes, we perform the translation v → v', effectively expressing the integrand in the special coordinate system. We take the u derivative inside the integral, observing that it can then be replaced by the u' derivative. We use equations (20) and (21) and their conjugates, integrating by parts where necessary, and then show that the terms proportional to A, A*, B and B* cancel pairwise.

**Conclusions**

In this paper we have studied a new formulation of a theory of fermions, which we take to be neutrinos, invariant under light-cone reflection. It uses a "minimal substitution" to construct a covariant derivative via the introduction of a scalar field subject to a constraint. By transforming to "special coordinates", in which the scalar field takes the role of one of the coordinates, we were able to find the solutions of the equations of motion, showing that there are 4 distinct choices for the mass-squared operator, two positive (tardyons) and two negative (tachyons). For each such choice there are two solutions, corresponding to a particle-anti-particle pair.

If any of this makes sense, there are many directions to carry the work forward, among them:

--are there other combinations of fields that realize LCR and perhaps provide alternative possibilities for the spectrum? Likewise, are there other differential operators that should be included in the Lagrangian? We are thinking of things like

$$i\varepsilon_{\mu\nu\sigma\lambda} n^\mu \gamma^\nu \gamma^\sigma D^\lambda = -2\gamma_5 (\gamma \cdot n)(\gamma \cdot D)\,,$$

which is LCR invariant and whose scaling under VSR could be absorbed by an appropriate factor of χ. Preliminary investigation of terms like these has revealed no new solutions of any interest, but this conclusion is not definitive;

--assuming the LCR-invariant fields describe neutrinos, can one construct appropriate interactions with the rest of the Standard Model? As remarked above, one probably will have to introduce a factor $n \cdot \gamma$ in the coupling in order to ensure that the unphysical modes proportional to $|b_1>$ and $|b_4>$ do not interact;

--can the theory be sensibly quantized? This will involve constructing a Hamiltonian, working out (anti)commutation relations, determining properties of the Hilbert space, etc.

--what is the significance of the "special" coordinate system? It is perhaps not surprising that ultimately the equations of motion reduce to two-dimensional ones, in order to allow for a symmetry between timelike and spacelike propagation. But the connection between the special system and the original coordinates, governed by h, has not been specified. What additional information, if any, can be brought to bear on the question of determining h? Presumably the original coordinate system is the one in which interactions with the rest of the universe should be formulated.